# Broken axial symmetry as essential feature to predict radiative capture in heavy nuclei

E. Grosse[a,*], A.R. Junghans[b], and R. Massarczyk[a,b]

[a] Institute of Nuclear and Particle Physics, Technische Universität Dresden, 01062 Dresden, Germany

[b] Institute of Radiation Physics, Helmholtz-Zentrum Dresden-Rossendorf, 01314 Dresden, Germany

ABSTRACT

Cross sections for neutron capture in the range of unresolved resonances are predicted for more than 140 spin-0 target nuclei with A>50. Allowing the breaking of spherical and axial symmetry in nearly all these nuclei a combined parameterization for both, level density and photon strength is obtained with surprisingly few fit parameters only. The strength functions used are based on a global fit to IVGDR shapes by the sum of three Lorentzians. They are based on theoretical predictions for the A-dependence of pole energies and spreading widths and add up to the TRK sum rule. For the small spins reached by capture resonance spacings are well described by a level density parameter close to the nuclear matter value; a significant collective enhancement is apparent due to the deviation from axial symmetry. Reliable predictions for compound nuclear reactions also outside the valley of stability – important for nuclear astrophysics and for the transmutation of nuclear waste – are expected to result from the global parameterization presented.

## 1. Introduction

The radiative capture of neutrons in the keV to MeV range by heavy nuclei plays an important role in considerations for advanced systems aiming for a reduction of radioactive nuclear waste [1]. This process is of interest also for the cosmic nucleosynthesis, especially for scenarios with neutron capture leading to a production of nuclides beyond Fe by the s-process [2]. Usually predictions for radiative neutron capture cross sections in the range of unresolved resonances are based on statistical model calculations. Their reliability depends not only on the proper characterization of the input channel, but more strongly on the details determining the decay of the intermediately formed compound nucleus. Here the strength of its electromagnetic decay is of importance as well as the open phase space in the final nucleus, i.e. the density of levels reached by the first photon emitted. The experimental studies forming the basis for parameterizations can mainly be performed on nuclei in or close to the valley of stability, but in cosmic environments many radiative processes occur in exotic nuclei which are not easily accessible experimentally. The knowledge of radiative neutron capture by unstable, *e.g.* actinide, nuclei is also of importance for the understanding of the competition between nuclear fission and the production of long-lived radionuclides by capture. It is thus desirable to derive a parameterization which is global and thus expected to be applicable also away from stable nuclei. It thus should rely on concepts valid generally and directly account for effects of nuclear shells and shapes. As is well known [3], the variation of nuclear quadrupole moments over the nuclide chart is very significant. It thus is indicated to investigate the influence of shape symmetries on nuclear level densities as well as on the extraction of photon strength functions.

The results of the various experiments on electromagnetic processes were previously often analysed [3] not regarding triaxiality. As demonstrated [4-7] Coulomb excitation studies have to carry out their data analysis considerably beyond the well documented [8] information on B(E2)-values and their relation to intrinsic quadrupole moments. Also theoretically the breaking of axial symmetry has often been disregarded, although it was shown [9] within the Hartree-Fock-Bogoliubov (HFB) scheme, that exact 3-dimensional angular momentum projection results in a triaxial minimum also for nuclei previously considered axially symmetric. Various spectroscopic studies [10, 11] have identified triaxiality effects in many nuclei. This is especially the case in nuclei with small quadrupole moments, but also seen in nuclei known to be well deformed (like actinides [7]). The low excitation level structure is dominated by the pairing degree of freedom, which induces Boson like modes, and triaxiality has been shown to be in good accord to interacting Bosons describing low energy phenomena in nuclear spectroscopy (IBA-2) [12]. In this work some use is made of a constrained CHFB-calculation for more than 1700 nuclei [13], which predicts not only quadrupole transitions rather well, but also the breaking of axial symmetry, *i.e.* the triaxiality parameter *γ*. Based on these results predictions were derived for the energy dependence of electric dipole strengths by a triaxial Lorentzian (TLO) parameterization of isovector giant dipole resonance (IVGDR) data. As recently demonstrated for nuclei with mass number A>70 [14-16] TLO yields good agreement to photo-neutron cross sections in the IVGDR with its deformation induced widening and splitting, and it also is in accord with electric dipole absorption data below the separation energy $S_n$. If the restriction to spherical or axial symmetry is released, the contribution of collective rotation to level densities increases these significantly [3, 17]. To allow for shape symmetry changes we will introduce a partly novel Fermi gas approach which explicitly considers triaxiality. Combining this collective enhancement with the dipole strength parameterization based on a global triple Lorentzian (TLO) fit to IVGDR data [14-16] a prediction is derived for radiative neutron capture in spin-0 target nuclei. As a test, a comparison to radiative capture cross sections in the energy range of 30 keV will be presented and a confirmation of the approach is obtained by regarding average resonance distances determined by neutron capture in 146 cases.

## 2. Level densities in nuclei without axial symmetry

To work out the effect of triaxiality on level densities we use an analytical approach for their calculation on an absolute scale. We avoid – as far as possible – parameter adjustments and strongly rely on statistical laws for a Fermi gas – a system of independent particles with mutual attraction. It is characterized by a gap $\Delta(t)$ falling with rising temperature *t* down to 0 at a 'critical' $t_{pt} = \Delta_o \cdot e^C/\pi = 0.567 \cdot \Delta_0$ [18-20] (with the Euler constant C=0.5772), indicating a 2$^{nd}$ order phase transition. Canonical thermodynamics is only used to evaluate the general features of this phase transition, but all effects appearing in finite nuclei will be treated micro-canonically. As discussed previously [17- 21], this may require various approximations. Albeit of minor influence for the conclusions made in this work they are listed already here (with proton and neutron number *Z* and *N*):

1. The pairing parameter $\Delta(E_x=0)$ is approximated by $\Delta_0 = 12 \cdot A^{-1/2}$, independent of angular momentum.
2. $\Delta_0$ is used for neutrons and protons and thus independent of neutron excess N-Z.
3. Quasi-particle states are evenly spaced (at least on average) at the Fermi energy, not varying with N-Z.
4. Fermi energy $\varepsilon_F$ =37 MeV and nuclear radius R= $r_0 \cdot A^{1/3}$ =1.2·$A^{1/3}$ are independent of N-Z.
5. A dependence of equilibrium deformation on excitation energy $E_x$ and angular momentum *J* is neglected.
6. The moments of inertia, which will be shown to have nearly no effect, are taken from a rigid rotor.

It is worth mentioning, that higher order dependences on $E_x$ and *J* are of minor importance for the comparison to experimental data taken from radiative neutron capture by spin-0-nuclei: The average resonance distances are observed in the region near $S_n \approx 7$ MeV and the state densities entering in the photon decay calculations and capture cross sections have to be known at $E_x \approx 3$ MeV, as discussed in Section 4. If only quasiparticle excitations are considered the total state density (in the intrinsic frame) $\omega_{qp}(E_x)$ at excitation energy $E_x$ is approximated by [21],

$$\omega_{qp}(E_x) = \omega_{qp}(0) \exp\left(\frac{E_x}{T_{ct}}\right) \quad \begin{array}{c}\text{for}\\ E_x < E_{pt}\end{array} \quad \text{and} \quad \omega_{qp}(E_x) = \frac{\exp(2\sqrt{\tilde{a}(E_x - E_{bs})})}{\frac{12}{\sqrt{\pi}} \tilde{a}^{1/4}(E_x - E_{bs})^{5/4}} \quad \begin{array}{c}\text{for}\\ E_x \geq E_{pt}\end{array} \quad (1).$$

At the phase transition energy $E_{pt} = \tilde{a} \cdot t_{pt}^2 + E_{bs}$ (corresponding to $t_{pt}$) a transition from a Fermi gas like behaviour above to a pairing dominated regime below the phase transition occurs. The latter is approximated by the assumption of constant temperature as will be discussed below together with the determination of the phenomenological parameters $T_{ct}$ and $\omega_{qp}(0)$ for this low energy region. In the Fermionic regime ($E \geq E_{pt}$), $E_{bs}$ stands for the backshift energy between the Fermi gas zero and the nuclear ground state and ã is the 'level density parameter'. In infinite nuclear matter (nm) the level density parameter is inversely proportional to the Fermi energy $\varepsilon_F$ and it determines the energy $E_{con}$ of the pairing induced condensation [18-20, 22, 23]:

$$a_{nm} = \frac{\pi^2 A}{4\varepsilon_F} \cong \frac{A}{15}; \quad \tilde{a} = a_{nm} + \delta a; \quad \delta a = \alpha \cdot A^{2/3} \text{ and } \quad E_{con} = \frac{3}{2\pi^2} a_{nm} \Delta_0^2; \quad E_{bs} = E_{con} - \delta E(Z, A) \quad (2).$$

It was shown [18] that the expression given under (1) for $\omega_{qp}$ in the Fermi gas regime – initially derived neglecting pairing [3, 21] – is a good approximation for the formalism derived with a thorough (micro-canonical) inclusion of pairing, if $E_x$ is back-shifted by the condensation energy $E_{con}$, which - in analogy with Fermionic systems in general - is independent of A. The back-shift $E_{bs}$ as given in Eq. (2) combines this pairing term with the effective shell correction $\delta E(Z,A)$ which is derived from nuclear ground state masses in comparison to liquid drop model calculations and which includes the odd-even mass difference. At variance to previous work [19, 20, 22, 24-26] it is subtracted from $E_x$ in Eq. (1) to directly correct for the energy lowering by shell effects in finite nuclei [21, 23, 27]. The influence of the nuclear surface is treated by changing $\tilde{a}$ by the global fit parameter $\alpha$ (actually the only one) quantifying the proportionality of $\delta a(A)$ to $A^{2/3}$. The intrinsic (quasi-particle) state density $\omega_{qp}(E_x)$ for the Fermionic region as well as for $E_{pt}$ are given by Eqs. (1) and (2). Below $E_{pt}$ an interpolation of $\omega_{qp}(E_x)$ to the ground state has to be found and the simple approach of a logarithmic interpolation in analogy with an exponential increase of $\omega_{qp}(E_x)$ has been shown to be a reasonable approximation to the low excitation structure of heavy nuclei [20-22, 24]. At variance to that work we use $\tilde{a}$, $t_{pt}$ and $E_{bs}$ to fix $E_{pt}$ and the requirement of a continuous transition in $\omega_{qp}(E_x)$ at $E_{pt}$ to determine $T_{ct}$ and $\omega_{qp}(0)$. The state density $\omega_{qp}(E_x<\Delta_0, J)$ at the lower end of the interpolation just above the ground state can be fixed here separately, as has been done *e.g.* in accordance with data previously [28-30]. In a first approximation we set it to $1/\Delta_0$ and we found a weak effect on the neutron capture cross section predictions.

From Eq. (1) one sees, that the Fermi-gas temperature parameter $t=\sqrt{(E_x-E_{bs})/\tilde{a}}$ (defined at the saddle point by approximating the Laplace transform [3, 18, 21, 22]) differs from an apparent nuclear temperature $T_{app} = \frac{\omega}{\partial\omega/\partial E}$. We find that also $T_{ct}$ is smaller than $T_{app}$ by up to 35% and this results in an equivalent sudden change in the slope of $\omega(E_x)$ at $E_{pt}$; near magic nuclei the large negative shell correction results in a different behaviour at the now large $E_{pt}$. In more than 100 of the 146 nuclei investigated here $E_{pt}$ is smaller than $S_n$ and thus the neutron capture resonances fall into the Fermi gas regime, but the subsequent gamma decay preferentially ends below $E_{pt}$ and the level density there dominates radiative capture cross sections. The quantities to be compared to observed level spacings have to be derived from $\omega_{qp}(E_x)$ by a projection on angular momentum $J$ in the observer system. The proposal was made [3, 21, 31, 32] to consider the M-substate distribution of $\omega_{qp}(E_x)$ as Gaussian with width $\sigma$ around $M = 0$ and to differentiate at $M = J+½$ with respect to M. This leads to a spin dependent level density [3, 17-22, 24-26]:

$$\rho_{sph}(E_x, J) \cong \frac{2J+1}{\sqrt{8\pi}\,\sigma^3} e^{-\frac{(J+½)^2}{2\sigma^2}} \omega_{qp}(E_x) \xrightarrow{small\,J} \frac{2J+1}{\sqrt{8\pi}\,\sigma^3} \omega_{qp}(E_x) \text{ with } \sigma = \sqrt{\frac{\Im t}{\hbar^2}} \qquad (3).$$

The spin dispersion $\sigma$ depends on the nucleus' moment of inertia $\Im$, often assumed to be the rigid rotor value [3]. The redistribution of the quasi-particle states into levels of distinct spin as incorporated here implicitly assumes [32] the nucleus to be exactly spherical symmetric even at $E_x = S_n$. This neglects strongly mixed modes which, due to their collectivity, are pulled from their original quasi-particle energy down into the low excitation regime. Albeit complete spherical symmetry was not assured, Eq. (3) has found a widespread use [18-22, 24- 26]. In a number of works, the rotational collectivity present in an axially symmetric nucleus was included at this stage [3, 17, 31, 32], yielding a level density enlargement by a factor $\sigma^2$ (i.e. $\approx A/5$) as compared to Eq. 3. But still an agreement with observations was not reached without a significant enlargement of $\tilde{a}$ as compared to $a_{nm}$ [3, 20, 30, 31].

This is why we did not use the scheme presented in Eq. (3) for the case of spherical symmetry, but selected to include the effect of missing axial symmetry for our comparison to experimental resonance spacings in 146 even-odd nuclei. As has been shown [3, 17] one then obtains – when considering a factor 1/4 for $\mathcal{R}$-symmetry conservation – for the density of levels with both parities:

$$\rho(E_x, J) \cong \frac{\sqrt{8\pi}}{4} \sigma_1\sigma_2\sigma_3 \frac{2J+1}{\sqrt{8\pi}\,\sigma^3} e^{-\Sigma_i \frac{(J+½)^2}{2\sigma_i^2}} \omega_{qp}(E_x) \xrightarrow{small\,J} \frac{2J+1}{4} \omega_{qp}(E_x) \qquad (4).$$

Allowing triaxiality already initially, Eq. (4) results from a summation "*over the different rotational levels in a given band having the same value of J*" [17]. The approximations used there to arrive at Eq. (4) analytically were tested by us numerically for various cases. The rotational energies $E_i = J \cdot (J+1) \cdot \hbar^2/\mathfrak{I}_i = J \cdot (J+1) \cdot t/\sigma_i^2$ – with i indicating the three body axes – have to be subtracted from $E_x - E_{bs}$ in the exponent in the numerator of Eq. (1) and the square root was expanded to obtain (in 2$^{nd}$ order) $\exp(-\sum E_i/t)$ as rotational energy correction, which finally leads to a spin cut off for each axis. Assuming an average equality of spin cut-off factors $\sigma_i$ and the spin dispersion $\sigma$ already appearing in Eq. (3) leads to a cancellation and thus to an independence on the moments of inertia $\mathfrak{I}$. Consequently Eq. (4) contains the $\sigma_i$ only in the exponential spin cut-off. As we limit ourselves to the case of s-capture by even nuclei into $J^\pi = \frac{1}{2}^+$ the influence of rotational energy and the corresponding cut off can be neglected here, and a surprisingly simple expression is obtained. For the limit of small $J$ it is presented in [17] and the text book of Bohr and Mottelson [3, Eq. (4-65b)]; future studies on the case of higher spins are needed to investigate the effect of the above approximations more thoroughly.

The inclusion of collectivity considerably increases the average level density at low energy by pulling quasiparticle states down into collective bands built on top of intrinsic parent states. With a typical spin dispersion (or cut off) factor of $\sigma \gtrsim 4$ an enhancement of more than A/2 results of $\rho(E_x, J)$ over $\rho_{sph}(E_x, J)$, which assumes conservation of sphericity (Eq. 3). As shown previously [3, 17] the enhancement is considerably reduced when axial symmetry is still assumed. It should be noted here that in previous studies [20-22, 31, 32] a rotational enhancement was treated as a correction to be applied only for nuclei assumed to conserve axial symmetry, whereas we allow a priori the breaking of that symmetry. This allows us to use the comparison to resonance spacings as a test of the nuclear symmetry at the respective energy. For a comparison to experimental level densities $\rho(E_x,J)$ of Eq. (4) has to be used; for compound nuclear reaction calculations (by Fermi's golden rule) the state density $\omega(E_x) = (2J+1) \cdot \rho(E_x,J)$ may be needed. Both differ considerably from the case of complete spherical symmetry, which has to be assured to set $\omega(E_x) = \omega_{qp}(E_x)$, as was often done in the past when combining Eq. (1) with Eq. (3). Doing so, *i.e.* assuming spherical symmetry *ad hoc*, an 'experimental' value $\tilde{a}_{exp}(A)$ was extracted for many A by using the average s-wave resonances spacings $D_{exp}(S_n, \frac{1}{2}^+) = 1/\rho_{exp}(E_x, \frac{1}{2}^+)$ as observed in neutron capture [20, 33]. By a few iterative steps account can be made for the appearance of $\tilde{a}^{\frac{1}{4}}$ in the denominator of Eq. (1). It has been demonstrated [20, cf. Figs. 24-26 and 29] for various modifications in the evaluation of $\omega_{qp}$ that this procedure always resulted in values close to $a_{exp} \gtrsim A/10$ in clear disagreement with $a_{nm} \cong A/15$. To test, if avoiding the *ad hoc* assumption of sphericity changes the situation, we have repeated such an extraction with the help of the 'triaxial' Eq. (4), using capture resonance data [20, 33]. The results are visualized in Fig.1 for two choices of $\delta E(Z,A)$ and without any shell correction; to simplify the situation, the calculations shown were done with $\delta a=0$. The overall agreement can be improved somewhat by setting $\alpha$ to 0.1 and 0.03 for the two choices for $\delta E$, respectively.

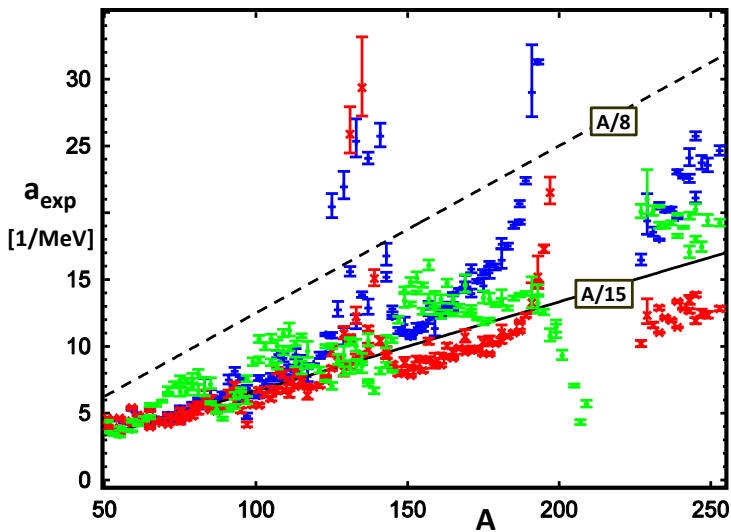

**Fig. 1:** Comparison of the apparent level density parameter $a_{exp}$ as extracted from resonance spacings observed for 51<A<253 to a linear dependence A/15 (drawn line) and A/8 (hatched line). Eqs. (1) and (4) were used to extract $a_{exp}$ from the data. Either no shell correction was applied (green ●), or the one from [34] (red **x**), respectively the one from [35], (blue **+**).

From Fig.1 three facts become obvious:
1. Most of the data for the 146 even-odd nuclei favour $a_{exp} \cong A/15$.
2. Rather strong deviations are observed near N=82 and near Z=82.
3. The choice of $\delta E$ (Z,A) has a significant effect, indicating that independent information is needed for a selection.

It should be stressed here, that our proposition to not exclude 'ad hoc' a deviation from spherical symmetry [17] combined with the consequent account for the condensation energy [18-20, 22, 23] clearly reduces the disagreement between $a_{exp}$ and the nuclear matter value $a_{nm} \cong A/15$. A similarly small $\tilde{a}$ was successfully used in a fragmentation study [23]. The contribution of vibrational collectivity was investigated as well on the basis of the respective expression from literature [17, 19, 20]: Inserting $\hbar\omega_{vib} = E_x(2^+,2)$ and $E_x(2^+,3)$ in with the energies $E_x(2^+)$ taken from the CHFB calculations [13] results in an enhancement of at most 35%, which we neglected in view of the large factor between Eqs.(3) and (4).

As was pointed out previously [20, cf. Fig. 22], various expressions derived from fits to ground state masses predict considerable differences for $\delta E(Z,A)$. In the following we will use the results of the fit presented in 1967 [35]; it yields results similar to more recent fits [24, 26]. We prefer it to an older one [34], which was favoured recently [20, 22, 25] for level density purposes. We use $\delta E(Z,A)$ as given for odd nuclei [35], i.e. without $\Delta_0$; as we concentrate on these, the comparison of experimental masses to this liquid drop prediction already contains the odd-even mass difference. Using $\alpha = 0.1$ the prediction for average resonance spacings is agreeing best with those observed in neutron capture. The result is shown in Fig.2 in comparison to neutron capture data for 146 even target nuclei with A>50 [20, 33]; the average spacing of s-wave resonances $D(S_n, J^\pi=½^+) = 1/\rho(E_x, J^\pi=½^+)$ is depicted. Our prediction is independent of the spin distribution, as all resonances have spin ½$^+$, and the small $J$ limit differs from the full expression with spin cut-off by a few % only.

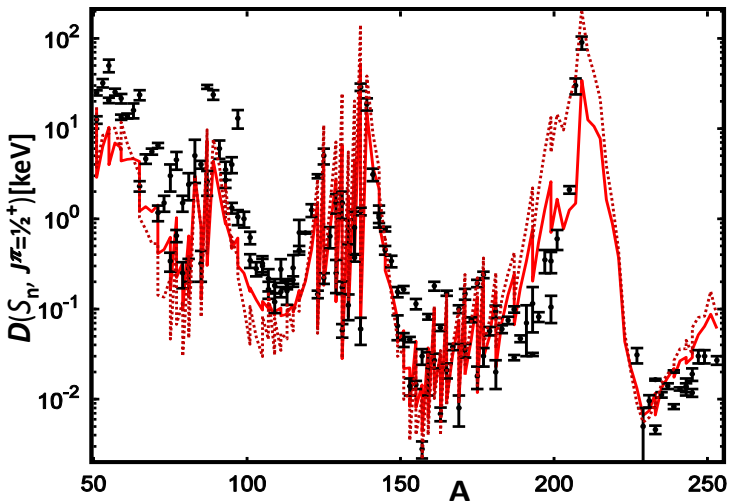

**Fig. 2:** Experimental information on average resonance spacings near $S$n ([33], black●) versus nuclear mass A. The prediction shown as drawn line (red) was obtained for $\alpha = 0.1$ including shell correction [35], damped with $E_x$ [27]; the dotted line depicts the no-damping case.

The figure shows the calculated level distances at $S_n$ including shell effects, either un-damped or with a damping related to the average frequency $\varpi_{sh}$ of the harmonic oscillator (determined by radius $R=r_0 \cdot A^{⅓}$ and nucleon mass $m_N$) [27]. Because of its small effect and its close relation to a well-established treatment for the energy dependence of shell structure [3] we omit details of this approximation. The observed difference to data for A≈208 is far below the factor of ≳100 mentioned above, indicating collective enhancement near $E_x \approx S_n$ even at closed shells.

In addition to capture resonances experimental level density information is available from spacings between bound nuclear levels as long as the observations have not missed any levels. The respective information is usually presented as apparent nuclear temperature $T_{app}^{exp}$ and – in view of scarce data – assumed to be independent of spin and parity. Because of the spin independence one can set $\rho(E) \propto \exp(E/T_{app}^{exp})$ and compare $T_{ct}(A)$ in Eq. (1) to $T_{app}^{exp}(A)$. This is done in Fig. 3 which depicts values extracted by various authors [20, 22, 24] from experimental information on nuclear level schemes and capture resonance spacings. The determination of $T_{ct}$ from our ansatz for the level density prediction is described above and obviously it characterizes the state distribution below the phase transition. In view of the scatter in the experimental values the agreement with the prediction is satisfactory. Also in Fig. 2 the measured data lie close to the prediction although only one free parameter was introduced, the small surface term α=0.1 in Eq. (2). This remarkable reduction of the number of free parameters is a clear advantage over previous proposals for analytic level density models [19, 20], which usually require at least four fit parameters without arriving at a more convincing agreement.

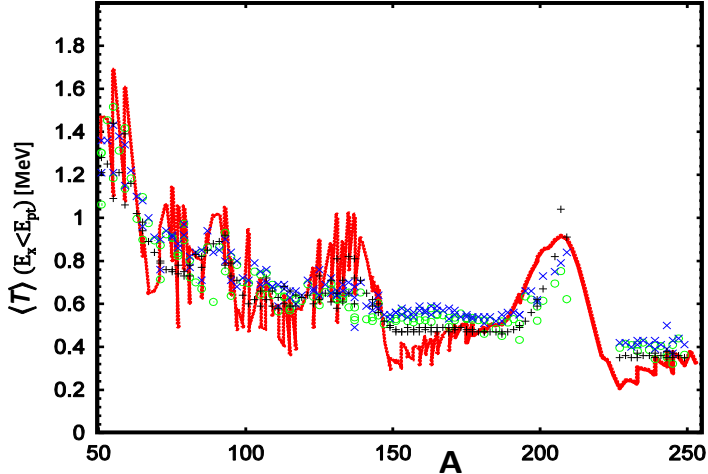

**Fig. 3:** Comparison of experimental information (green ○ [20], black ✚ [22], blue ✖ [24]) on the apparent nuclear temperature $T_{app}^{exp}$ (in MeV) to results predicted for $E_x<E_{pt}$ depicted as a red line.

Obviously the new finding of triaxiality being a very common property of excited heavy nuclei considerably affects our understanding of the nuclear level statistics. Another important influence on ρ($E_x$,$J$) results from the choice made for the shell correction δ$E$: It is not serious, but the effect increases with A, indicating the need of further theoretical study. In any case the new level density formalism should be applied to the analysis of compound nuclear cross section data. Using information on photon strength as presented in the next Section this is done for radiative neutron capture in Section 4.

## 3. Dipole strength in triaxial nuclei – including odd isotopes.

Electromagnetic processes play an important role not only in nuclear spectroscopy but also for the de-excitation processes following neutron capture or other nuclear reactions. Since decades the relation of the IVGDR to the nuclear radiative strength [36, 37] is considered the basis of its parameterization for heavier nuclei. Its mean position $E_0$ can be predicted using information from droplet model fits to ground state masses and a parameterization of the electromagnetic strength in heavy nuclei with mass number A>70, which considers their triaxial deformation, was shown [14] to be in reasonable accordance with measurements. For triaxial nuclei the three pole energies $E_k$ are given by the three axis lengths $r_k$ : $E_k$ = $r_0$/$r_k$·$E_0$ obtained from *a priori* information on the deformation. Using averages from the even neighbours this is the case also for odd target nuclei and Eq. (5) describes electric dipole strengths $f_{E1}(E_γ)$ for both cases [37, 38] (with the fine structure constant $α_e$ and the nucleon mass $m_N$):

$$f_{E1}(E_γ) = \frac{\langle σ_{abs}(E_γ)\rangle}{(π\hbar c)^2 \, g_{eff} \cdot E_γ} = \frac{4\, α_e}{3π \, g_{eff}} \frac{ZN}{m_N c^2 \, A} \sum_k \frac{E_γ Γ_k}{\left(E_k^2-E_γ^2\right)^2+E_γ^2 Γ_k^2}; \quad g_{eff} = \sum_{J_r} \frac{2J_r+1}{2J_b+1} = 2λ+1 \quad (5).$$

The resulting triple Lorentzian (TLO) approach [15, 16] reproduces the mean absorption cross section $\langle σ_{abs}(E_γ)\rangle$ in the IVGDR region as well as in the low energy tail at energies below $S_n$ [14-16]. Here the relation between GDR pole energies $E_k$ and widths $Γ_k$, related by hydrodynamics, was generalized for triaxial shapes [39]: $Γ_k$ = $c_w$ ·$E_k^{1.6}$. The cross section $\langle σ_{abs}(E_γ)\rangle$ averaged over many compound nucleus (mini-)resonances with spin $J_r$ is directly observed by absorption of a photon spectrum containing the IVGDR by the target ground state $J_b$. Together they form the 'giant' IVDR and the energy integrated cross section is proportional to N·Z/A, as predicted [40] from fundamental considerations (TRK sum rule), summed over three components corresponding to the three axes. The sum for $g_{eff}$ runs over the resonance spins $J_r$ which can be formed from $J_b$ and the multipolarity $\lambda$. For two nuclei the TLO sum for the IVGDR is compared in Fig. 4 to rescaled data [41]; the three poles corresponding to axis ratios from CHFB are indicated as black bars. In previous work [14] we have shown how the axis ratios relate to the deformation parameters; using spectroscopic information a satisfying agreement with observed GDR shapes was reached with a proportionality factor $c_w \cong 0.05$ (in MeV units). When the CHFB calculations [13, cf. Eq.(3)] became available for 1634 nuclei with 50<A we decided to use these results on β, γ and the corresponding axis ratios to dispose of values for exotic nuclei; a predicted reduction [42] of the deformation parameter β near closed shells was adjusted to 2.3 from a comparison to data. With these the analysis [14] of photonuclear data for more than 50 isotopes arrived at a reasonable fit by using $c_w = 0.045$; the small difference to the 'old' value is of negligible importance for the calculations presented here. Clearly the data as shown in Fig. 4 are reproduced in accord with Eq. (5) – for both nuclei, although these have often been considered spherical. This supports the validity of our TLO approach together with the TRK sum rule. Lorentzian fits [20, 43] performed under the assumption of only one or two poles for the IVGDR clearly exceed the TRK sum rule, and their difference to TLO increases with decreasing photon energy [14]. This feature is significant for radiative capture which populates an excitation energy region at $E_x \lesssim S_n$ of high level density $\rho(E_x)$ with small $E_\gamma = S_n - E_x$. At such small energy $f_{E1}(E_\gamma)$ is determined in TLO predominantly by the width parameter. As already shown previously [14-16, 41], the TLO prediction is close to or below experimental data acquired by photon scattering or other radiative processes [37, 41], and experimental evidence is missing which would imply a need for an energy-dependent strength reduction proposed on the basis of IVGDR fits neglecting triaxiality [20, 43]. The agreement for both nuclei with Eq. (5) on absolute scale is a manifestation of the previously stated independence [37] of the photon strength $f_\lambda(E_\gamma)$ on the spins: $J_r$ and $J_b$ are replaced by $\lambda$. By adoption of the Axel-Brink hypothesis [36] we generalize to non-zero $E_b$, assume absorption and emission of photons to be described by the same $f_\lambda(E_\gamma)$ and obtain photon widths (averaged over $E_r$) for the dipole ($\lambda=1$) decay from resonances $J_r$ to bound states $J_b$ [37].

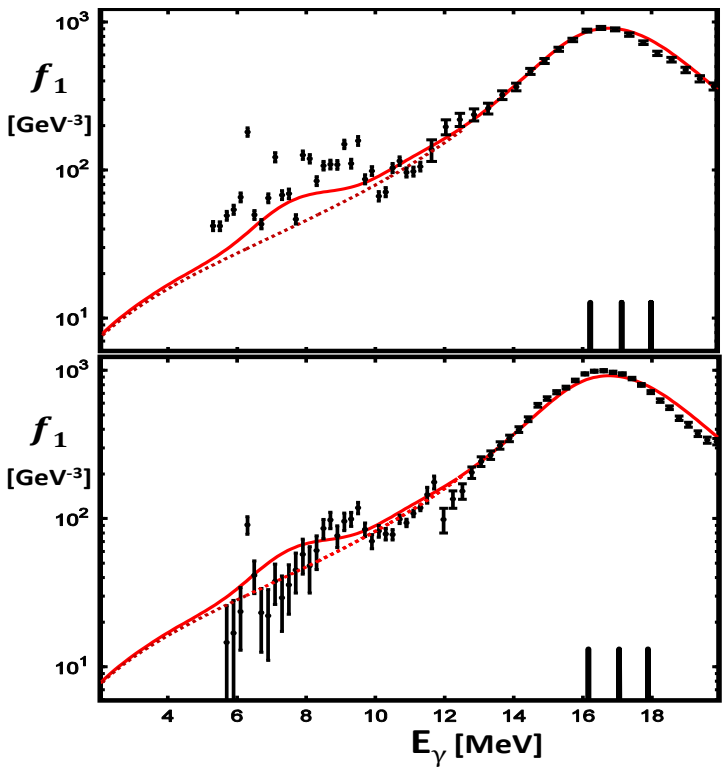

Fig. 4: Comparison of our parametrization to experimental data on photonuclear processes in $^{88}$Sr (top) and $^{89}$Y (bottom). Data above 12 MeV are from (γ,n) [44] and the others were obtained from photon scattering [45, 46]. A dotted line depicts the E1-strength predicted by Eq. (5) with the poles indicated as bars; the results obtained with the inclusion of minor strength are presented as full (red) line. The strong strength excess near 6 MeV may be related to a strong single particle excitation.

A sum over the decay channels to all bound states $J_b$ which can be reached by dipole photons of energy $E_\gamma = E_r - E_b$ from the resonances $J_r$, populated by capturing the neutron, leads to a second averaging, indicated in Eq. 6 by $\bar{\Gamma}_\gamma$.

The dependence of ρ($E_b$,$J_b$) on $J_b$ (cf. Eq. (4)) and the quantum-statistics for the number of magnetic sub-states of $J_b$ reached by the γ-decay have two consequences: for λ=1-transitions from $J_r$=1/2 to $J_b$=1/2 and $J_b$=3/2 it leads to a weight factor of *g* = 5 which then replaces the sum in Eq. (6). A difference for E1 and M1 only arises, if there is a parity dependence of the level density. As we will show the E1 decay to be predominant, this is of minor importance; in principle we can account for differences in ρ($E_x$,J) in the low energy regime (*e.g.* due to parity) by a respective estimate of ω($E_x$<$\Delta_0$) and we can improve our predictions, if respective information requires so. The mean radiative width is the basis for the description of radiative capture as discussed in Section 4 for odd final nuclei. Here the extrapolation of the nuclear electric dipole strength to $S_n$ and below – i.e. the low energy tail as given by Eq. (5) – is of importance. It was pointed out previously [43] that strength information can be extracted from capture data directly by regarding average radiative widths $\langle \bar{\Gamma}_\gamma \rangle_r$. Eq. (6) shows, that these are proportional to the photon strength, and depend in addition on the ratio between the level densities at the capturing resonances *r*- included in $f_1(E_\gamma)$ - and the final states *b* reached by the γ-decay. Consequently the average radiative widths vary with the slope of ρ($E_x$) in the range from $E_b$ to $E_r$, whereas capture cross sections also vary with the level density at $S_n$. A good agreement was found [15] between average radiative widths as derived by a resonance analysis of neutron data taken just above $S_n$ and tabulated [33] for over 120 even-odd nuclei (A>50) and $\langle \bar{\Gamma}_\gamma \rangle_r$ from Eq. (6) and TLO – with minor strength, as described in the following, included.

$$\langle \Gamma_\gamma(E_\gamma, J_b \leftrightarrow J_r) \rangle_r = \frac{f_1(E_\gamma) E_\gamma^3}{\rho(E_r, J_r)}; \quad \langle \bar{\Gamma}_\gamma \rangle_r \equiv \langle \langle \Gamma_\gamma(E_r, E_b) \rangle_b \rangle_r = \sum_{J_b} g \int_0^{E_r} \frac{f_1(E_\gamma) E_\gamma^3}{\rho(E_r, J_r)} \rho(E_b, J_b) \, dE_\gamma \quad (6).$$

At low energies photon strength components, which are not of isovector electric dipole character, contribute to radiative capture [20, 28-30, 43, 47-50] and our analysis aims for a rough estimate of their importance. Respective information from photon scattering [38, 51-53] is of use, asserting equal integrated strength for collective modes based on nuclear ground states and those on top of excited states [36, 37]. Minor strength, partly of M1 type, may also be derived from the analysis of gamma-decay following nuclear reactions [54-57]. Three such components, as apparent in Fig. 4 (two depending on the deformation β), have some impact on the predictions for radiative capture, as later shown in Section 4:

1. Orbital magnetic dipole strength (scissors mode [48, 52, 57]), which is approximated to peak at
   $E_{sc}$ = 0.21·$E_0$ with a maximum of $f_1^{max}$ = $Z^2$·$β^2$/76 GeV$^{-3}$, Gaussian distributed with σ = 1 MeV.
2. Electric dipole strength originating from coupled $2^+$ and $3^-$-phonons [51] is assumed to peak
   around $E_{quad}$ + $E_{oct}$ = $E_{qo}$ ≈ 3 MeV with a maximum of $f_1^{max}$ = Z·A·β/250 GeV$^{-3}$.
3. Electric dipole strength at $E_{py}$ ≈ 0.4$E_0$ - 0.5$E_0$ – known as pigmy mode [37] – observed in many nuclei to also show up in isoscalar processes [55, 56], recently reviewed [49] to approximately add 12 GeV$^{-3}$ to TLO.

Also for 2 and 3 a Gaussian distribution with σ = 1 MeV is assumed, as no fundamental reasons are given for a Lorentzian shape [43, 47]. It is admitted, that the guesses as presented here can only serve as a very first hint on the eventual role of these strength components. The magnetic strength related to nucleon spin-flip modes [43, 47, 52] appears at energies near $S_n$ and can thus be neglected in the discussion of radiative neutron capture mainly invoking photons of considerably smaller energy.

## 4. Radiative neutron capture

The good agreement of the low energy slopes of the IVGDR with a 'triple Lorentzian' parameterization (TLO) as obtained by using independent information on triaxial nuclear deformation suggests the use of a corresponding photon strength function also for the radiative neutron capture, an electromagnetic processes alike, combined with an expression for level densities valid in the case of reduced symmetry. To test the influence of dipole strength functions on radiative neutron capture over a wide range in A the investigation of only s-wave capture by spin 0 target nuclei has the advantage of offering a large sample with the same resonance spin and parity ½$^+$, and they decay by E1 to $1/2^-$ or $3/2^-$. As known from measured neutron strengths [33] the neutron widths above 5 keV are that large, that $\langle \Gamma_n \rangle_r \gg \langle \bar{\Gamma}_\gamma \rangle_r$ and the average over the width ratio can be replaced by $\langle \bar{\Gamma}_\gamma \rangle_r$ as given in Eq. (6). Porter-Thomas effects [36, 37] were approximated by using a factor of 0.8, derived from calculating statistical averages over a large number of neutron resonances $r$, and thus we arrive at $g' \cong 4$, which results in Eq. (7) for the radiative capture [58] (neglecting $\ell > 0$, direct capture and inelastic scattering):

$$\langle \sigma(n,\gamma) \rangle_r \cong 2\pi^2 \lambdabar_n^2 \, \rho(E_r, 1/2^+) \, \langle \frac{\Gamma_n \cdot \bar{\Gamma}_\gamma}{\Gamma_n + \bar{\Gamma}_\gamma} \rangle_r \cong 2\pi^2 \, \lambdabar_n^2 \cdot \sum_{J_b} g' \int_0^{E_r} f_1(E_\gamma) E_\gamma^3 \cdot \rho(E_b, J_b) dE_\gamma \qquad (7).$$

Covering the full range of A>50 in the comparison to data Maxwellian averaged (MACS) neutron capture cross sections are shown in Fig. 5 together with the prediction made by folding of the cross sections as given by Eq. (7) with a Maxwellian distribution of neutron energies [2]. MACS have been tabulated [59] covering many heavy nuclei as they are of use for the investigation of nuclear processes in cosmic objects like red giant (AGB) stars, where radiative neutron capture takes place at approximately $kT_{AGB}$ = 30 keV. For several actinide nuclei equivalent data were compiled [60] and uncertainty bars were derived from the scatter as published. In view of the fact that $D \gg \Gamma_R \geq \Gamma_{R\gamma}$ the Maxwellian averages around 30 keV are formed incoherently and fluctuations (beyond the ones mentioned above) are neglected. The good agreement on an absolute scale with data as displayed in Fig. 5 gives a convincing impression for the validity of the parameterization presented and the approximations applied.

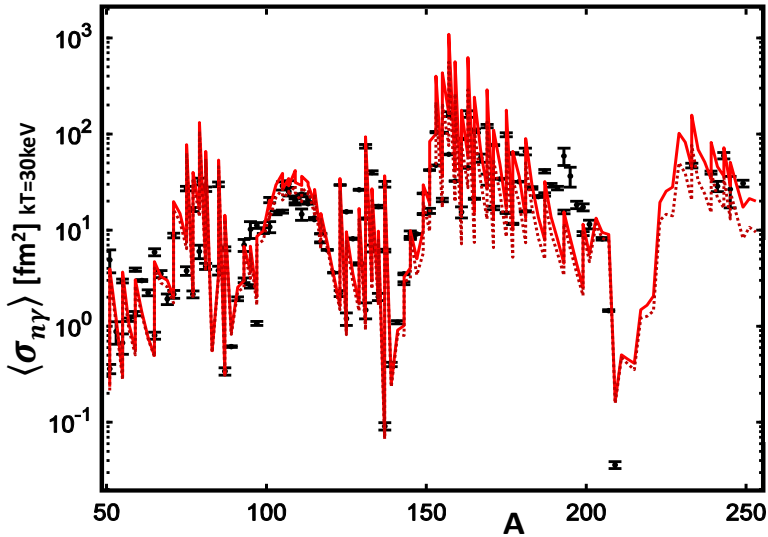

**Fig. 5:** Comparison of predicted neutron capture cross sections $\langle \sigma(n,\gamma) \rangle_r$ (full red curve, $\ell=0$) to experimental data on Maxwellian averaged cross sections [59] for $kT_{AGB}$ = 30 keV vs. A. The dotted curve was calculated with TLO only.

By regarding the radiative capture by spin-zero targets effects related to ambiguities of spin cut off parameter and angular momentum coupling are suppressed, but still the data vary by about 4 orders of magnitude in the discussed range of A. They are well represented by the TLO-parameterization, when minor photon strength as discussed at the end of Section 3 is included here and when the schematic ansatz for $\rho(A, E_x)$, as described by Eqs. (1, 2, 4 and 5), is used. Discrepancies appearing for some A may well be related to our omission of p-capture, which is known to be non-negligible in some mass range [20, 28]. This and other local effects have minor significance on the importance of broken axial symmetry in excited heavy nuclei – the main topic here.

Neutron capture by actinide nuclei is of great importance for the transmutation of nuclear waste and we investigate neutron capture cross sections for Th, U and heavier nuclei for which data [50,60] exist. Whereas the approximations made to arrive at Eq. (7) work well for $E_n \approx 30$ keV (see Fig. 5) the coupling to other channels like inelastic scattering has to be included. This may increase the calculated cross section, especially at higher neutron energy, as was shown [50, 57] in detail recently for $^{238}$U, where also the importance of the scissors mode was pointed out.

## 5. Conclusions

Various spectroscopic data indicate triaxiality for a number of heavy nuclei [4-6, 7, 10, 11]; two effects – hitherto not emphasised as such – indicate for nearly all of them a breaking of axial symmetry at higher excitation:

1) With one global parameter the scheme proposed here reproduces observations for level densities in nuclei with A > 50 and J = ½, when (a) the condensation energy $E_{con}$ is included in the Fermi gas backshift and (b) the collective enhancement due to symmetry reduction by triaxiality is included. This is achieved although the level density parameter ã has to be modified little from its nuclear matter value to fit resonance spacing data: The free surface correction term is rather small – much smaller than usual [20].

2) Again only one global parameter suffices to fit to the shape of the IVGDR peak by a triple Lorentzian photon strength (TLO) – considerably improved and in accord with the TRK sum rule. It also predicts its low energy tail – without other modifications than the addition of minor modes – to match respective strength data as well as neutron capture cross sections taken in the energy range of unresolved resonances.

For the last-mentioned finding a combination of the points 1) and 2) is needed, which is easily performed by considering spherical and axial symmetry to be broken – as shown for low excitation by HFB calculations [9, 13] and as expected to increase with energy. Exact deformation parameters are unimportant for the tail of the E1-resonance as well as for the density of low spin states occurring in neutron capture by even targets as neither spin cut off nor moments of inertia are involved. At variance to previous work [*e.g.* 20] the breaking of axial symmetry in excited heavy nuclei is demonstrated here on the basis of experimental data: For more than 140 spin-0 target nuclei with A>50 level distance data and average capture cross sections are well predicted by a global ansatz.

Within this work a literature study indicates a non-negligible effect of 'minor' magnetic and electric dipole strength (other than isovector electric): Photon data in the region of $E_\gamma$=3-5 MeV indicate that such strength may increase the radiative capture cross section by up to 100%. The global parameterization proposed here for isovector strength (TLO) with these additions agrees well to radiative neutron capture cross sections [59, 60] as shown in Fig. 5. As it also does not exceed directly measured photon strength in the region below $S_n$ [14-16, 41] it can be considered as good ingredient for network calculations in the field of cosmic element production as well as for simulations of nuclear power systems and the transmutation of radioactive waste, were predictions for actinide nuclei are of importance. Previous studies in the field of photon strength [*e.g.* 20, 43, 47, 57] have worked with a lower IVGDR tail leading to a larger relative influence of 'minor' strength components. Here the often assumed reduction of the resonance width with decreasing $E_\gamma$ plays an important role and the modifications [43, 47] added to increase $f_{E1}$ at small energies without much of a change in the peak region lead to a questionable prediction for 3-5 MeV. Similarly single or 2-pole IVGDR fits [16] are likely to create incorrect estimates of the relevant E1-strength as they result in an irregular A-dependence of the spreading width $\Gamma_{E1}$ and the resonant cross section integral in disagreement with the TRK sum rule. This sheds some doubt on E1 strength predictions presented by RIPL [20] which obviously lead to such irregularities. In contrast the triple Lorentzian scheme (TLO) with a variation of $\Gamma_{E1}$ with the pole energy $E_0$ alone uses only one global parameter (the proportionality between $\Gamma_{E1}$ and $E_0$) and accords to the TRK sum rule resulting in a global dipole strength prediction for the tail region. The ansatz presented here assumes the breaking of spherical or axial symmetry for nearly all heavy nuclei – at least near $S_n$ and above, where the resulting collective enhancement improves the description of resonance spacing data – also using only one global parameter. The remarkable reduction of fit parameters for level density and photon strength increases the predictive power for radiative capture processes. It is thus of interest to apply it in calculations for other compound nuclear reaction rates. Regarding the rather limited theoretical work done so far [4, 9, 12, 13, 39] the importance of broken axial symmetry already at low spin – as advocated here – should induce further investigations.

**Acknowledgements**
This work was supported by the project TRAKULA funded by the German Federal Ministry for Education and Research (contract number 02NUK013A) and the project ERINDA (contract numberFP7-269499) funded by the European Commission; R. M. had support from DFG, Contract No.SCHW883/1-1.
Discussions with H. Feldmeier, K.-H. Schmidt, R. Schwengner and H. Wolter are gratefully acknowledged.

**References**

1. M. Salvatores and G. Palmiotti, Prog. Part. Nucl. Phys. 66 (2011) 144.
2. F. Käppeler et al., Rev. Mod. Phys. 83 (2011) 157.
3. A. Bohr and B. Mottelson, Nuclear Structure ch. 2, 4 & 6, (Benjamin, Reading, Mass., 1975).
4. K. Kumar, Phys. Rev. Lett. 28 (1972) 249.
5. J. Stachel et al., Nucl. Phys. A383 (1982) 425
6. D. Cline, Ann. Rev. Nucl. Part. Sci. 36 (1986) 683.
7. C. Y. Wu and D. Cline, Phys. Rev. C 54 (1996) 2356.
8. S. Raman et al., At. Data and Nucl. Data Tables 78 (2001) 1.
9. A. Hayashi, K. Hara and P. Ring, Phys. Rev. Lett. 53 (1984) 337.
10. W. Andrejtscheff and P. Petkov, Phys. Rev. C 48 (1993) 2531; id., Phys. Lett. B 329 (1994)1.
11. Y. Toh et al., Phys. Rev. C 87 (2013) 041304.
12. F. Iachello, Phys. Rev. Lett. 91 (2003) 132502.
13. J.-P. Delaroche et al., Phys. Rev. C 81 (2010) 014303; id., supplemental material.
14. A.R. Junghans et al., Phys. Lett. B 670 (2008) 200.
15. R. Beyer et al., Int. Journ. of Mod. Phys. E20 (2011) 431.
16. A.R. Junghans et al., Journ. Korean Phys. Soc. 59 (2011) 1872.
17. S. Bjørnholm, A. Bohr and B. Mottelson, Rochester-conf., IAEA-STI/PUB/347 (1974) 367.
18. M.K. Grossjean and H. Feldmeier, Nucl. Phys. A 444 (1985) 113.
19. A.V. Ignatyuk et al., Phys. Rev. C 47 (1993) 1504; id., IAEA-INDC 0233 (1985) 40.
20. R. Capote et al., Nucl. Data Sheets 110 (2009) 3107; id., //www-nds.iaea.org/RIPL-3/
21. A. Gilbert and A.G.W. Cameron, Can. Journ. of Phys. 43 (1965)1446
22. A. Koning et al., Nucl. Phys. A 810 (2008) 13
23. A.R. Junghans et al., Nucl. Phys. A 629 (1998) 635.
24. T. v. Egidy and D. Bucurescu, Phys. Rev. C 72 (2005) 044311.
25. A. Mengoni and Y. Nakajima, J. Nucl. Sci. & Technol. 31(1994) 151.
26. S.F. Mughabghab and C. Dunford, Phys. Rev. Lett. 81 (1998) 4083.
27. S.K. Kataria, V. S. Rarnamurthy, and S. S. Kapoor, Phys. Rev. C 18 (1978) 549.
28. G. Schramm et al., Phys. Rev. C 85 (2011) 014311.
29. G. Rusev et al., Phys. Rev. C 87 (2013) 054603; id., Phys. Rev. Lett. 110 (2013) 022503.
30. R. Massarczyk et al., Phys. Rev. C 86 (2012) 014319; id., Phys. Rev. C 87 (2013) 044306;
31. J. R. Huizenga et al., Nucl. Phys. A 223 (1974) 589.
32. S.E. Vigdor and H.J. Karwowski, Phys. Rev. C 26 (1982) 1068.
33. A.V. Ignatyuk, RIPL-2, IAEA-TECDOC-1506 (2006); www-nds.iaea.org/RIPL-3/resonances
34. W.D. Myers and W.J. Swiatecki, Nucl. Phys. 81 (1966) 1.
35. W.D. Myers and W.J. Swiatecki, Ark. Fizik 36 (1967) 343.
36. P. Axel, Phys. Rev. 126 (1962) 671; quoted therein: D. Brink, Ph.D. thesis, Oxford 1955.
37. G. A. Bartholomew et al., Adv. Nucl. Phys. 7 (1972) 229.
38. E. Grosse and A.R. Junghans, Landolt-Börnstein, New Series I, 25 (2013) 4.
39. B. Bush and Y. Alhassid, Nucl. Phys. A 531 (1991) 27.
40. M. Gell-Mann et al., Phys. Rev. 95 (1954) 1612.
41. M. Erhard et al, Eur. Phys. Journ. A 27 (2006) 135; id., Phys. Rev. C 81 (2010) 034319.

42  G. F. Bertsch et al., Phys. Rev. Lett. 99 (2007) 032502.
43  J. Kopecky and M. Uhl, Phys. Rev. C 41 (1990) 1941.
44  A. Lepretre et al., Nucl. Phys. A 175 (1971) 609.
45  R. Schwengner et al., Phys. Rev. C 76 (2007) 034321.
46  N. Benouaret et al., Phys. Rev. C 79 (2009) 014303.
47  J. Kopecky, M. Uhl and R.E. Chrien, Phys. Rev. C 47 (1993) 312.
48  M. Krticka et al., Phys. Rev. Lett. 92 (2004) 172501.
49  R. Massarczyk et al., Phys. Rev. Lett. 112 (2014) 072501.
50  J.L. Ullmann et al., Phys. Rev. C 89 (2014) 034603.
51  U. Kneissl et al., J. Phys. G 32 (2006) R217; id., Nuclear Physics News 16 (2006) 27.
52  K. Heyde et al., Rev. Mod. Phys. 82 (2010) 2365.
53  E. Kwan et al., Phys. Rev. C 83 (2011) 041601
54  NNDC database: https://www.nndc.bnl.gov/exfor/exfor.htm
55  T.D. Poelhekken et al., Phys. Lett. B 278 (1992) 423.
56  D. Savran et al., Phys. Rev. Lett. 100 (2008) 232501.
57  M. Guttormsen et al., Phys. Rev. Lett. 109 (2012) 162503; id., Phys. Rev. C 88 (2013) 024307.
58  A. M. Lane and J. E. Lynn, Proc. Phys. Soc. (London) A70 (1957) 557.
59  I. Dillmann et al., Phys. Rev. C 81, 015801 (2010); id., //www.kadonis.org.
60  B. Pritychenko et al., At. Data and Nucl. Data Tables 96 (2010) 645; www.nndc.bnl.gov/astro.